\documentclass[adp,fleqn]{w-art}
\usepackage{times}
\usepackage{cite}
\usepackage[]{graphicx}
\begin{document}
\DOIsuffix{/DOI 10.1002/andp.201100299}
\Volume{524}
\Issue{8}
\Month{01}
\Year{2007}
\pagespan{3}{}
\Receiveddate{9 November 2011}
\Reviseddate{7 July 2012}
\Accepteddate{9 July 2012}
\Dateposted{23 July 2012}



\title[Event-by-event simulation of quantum phenomena]{Event-by-event simulation of quantum phenomena}


\author[H. De Raedt]{H. De Raedt\inst{1}%
  \footnote{Corresponding author\quad E-mail:~\textsf{h.a.de.raedt@rug.nl},
            Phone: +31\,50\,3634950,
            Fax: +31\,50\,3634947}}
\address[\inst{1}]{
Department of Applied Physics, Zernike Institute for Advanced Materials,
University of Groningen, \\Nijenborgh 4, NL-9747 AG Groningen, The Netherlands
}
\author[K. Michielsen]{K. Michielsen\inst{2,3}\footnote{E-mail:~\textsf{k.michielsen@fz-juelich.de}}}
\address[\inst{2}]{
Institute for Advanced Simulation, J\"ulich Supercomputing Centre,
Forschungszentrum J\"ulich, D-52425 J\"ulich, Germany
}
\address[\inst{3}]{
RWTH Aachen University, D-52056 Aachen, Germany}
\begin{abstract}
A discrete-event simulation approach is reviewed that does not require the knowledge of the solution of
the wave equation of the whole system, yet reproduces the statistical distributions of wave theory by
generating detection events one-by-one.
The simulation approach is illustrated by applications to a two-beam interference experiment
and two Bell test experiments, an Einstein-Podolsky-Rosen-Bohm experiment with single photons
employing post-selection for pair identification
and a single-neutron Bell test interferometry experiment with nearly 100\% detection efficiency.
\end{abstract}


\maketitle                   






\def\DLM{DLM}
\def\DLMS{DLMs}
\def\Eq#1{(\ref{#1})}
\newcommand\sumprime{\mathop{{\sum}'}}
\newcommand\MT{Maxwell's theory}
\newcommand\QT{quantum theory}
\newcommand\url[1]{{\sl #1}}
\newcommand\ns{\,\mathrm{ns}}
\newcommand\mus{\,\mu\mathrm{s}}
\newcommand\EBM{event-based model}
\newcommand\EBMs{event-based models}

\section{Introduction}
\subsection{Quantum theory and the observation of single events}
Quantum theory has proven to be extraordinarily powerful for describing the
statistical properties of a vast number of laboratory experiments.
Starting from the axioms of quantum theory it is, at least conceptually, straightforward
to calculate numbers that can be compared with experimental data
as long as these numbers refer to statistical averages.
However, if an experiment records individual clicks of a detector a fundamental problem appears.
Although quantum theory provides a recipe to compute the frequencies for observing events
it does not account for the observation of the individual events themselves,
a manifestation of the quantum measurement problem~\cite{HOME97,BALL03}.
A recent review of various approaches to the quantum measurement problem and an explanation of it within
the statistical interpretation is given in Ref.~\cite{NIEU11b}.
From the viewpoint of quantum
theory, the central issue is how it can be that experiments yield
definite answers. As stated by Leggett~\cite{LEGG87}, ``In the final analysis,
physics cannot forever refuse to give an account of how it
is that we obtain definite results whenever we do a particular
measurement''.
In fact, a common feature of all probabilistic models
is that they do not entail a procedure that specifies
how particular values of the random variables are being realized, but
only contain a specification of the probabilities with which these values appear.
As a probabilistic theory, quantum theory takes a special position
in that it postulates that it is fundamentally impossible to go beyond
the description in terms of probability distributions.
At the present time, there is no scientific evidence
that supports this assumption other than that it seems unsurmountable to account
for the observation of the individual events
within the context of quantum theory proper~\cite{HOME97,BALL03}.
This suggests that the search for a cause-and-effect description of the observed phenomena
should be done outside the realm of quantum theory.

\subsection{From single events to probability distributions and not vice versa}
In general, the problem can be posed as follows:
Given a probability distribution of observing events, can we construct an
algorithm which runs on a digital computer and produces events
with frequencies that agree with
the given distribution {\bf without} the algorithm referring,
in any way, to the probability distribution itself.
An affirmative answer to this primary question can give rise to secondary questions such as
``Can this be done efficiently?'', ``Is the algorithm unique?'', and so on.

There are several elements in the statement-of-the-problem that deserve attention.
The first is that we use ``event'' in the every-day meaning of the word.
Thus, an event can be a click of a detector, the experimenter pushing a button, and so on.
Second, a digital computer is nothing but a macroscopic physical device that
evolves in time by changing its state in a well-defined,
cause-and-effect, discrete-event manner, as specified by the algorithm.
Although not practical, it is possible to build a mechanical
apparatus that performs exactly the same function as the digital computer.
Thus, one can view a simulation on a digital computer
as a perfectly controlled experiment on a macroscopic mechanical system,
simulating, in a cause-and-effect manner,
the phenomenon that is being observed in the laboratory.
Third, and most importantly, the algorithm
should not rely in any way on the probability distribution of the events that it is supposed to generate.
Otherwise it would be straightforward to
use pseudo-random numbers and generate events according to this distribution.
However, this is not a solution to the posed problem as it
assumes that the probability distribution of the quantum mechanical problem is known, which
is exactly the knowledge that we want to generate without making reference to quantum theory.
Put differently, the algorithm should be capable of generating events
according to an unknown probability distribution.

In order to clarify our aim we draw an analogy~\cite{RAED07c} with the Metropolis Monte Carlo method
for simulating classical statistical mechanics~\cite{LAND00}, a primary example of a method that samples from an unknown probability distribution.
According to the theory of equilibrium statistical mechanics, the probability that a system is in the state with label $n$ is given by
$p_n=e^{-\beta E_n}/\sum_{n=1}^Ne^{-\beta E_n}\equiv e^{-\beta E_n}/Z$, where $N$ is the number of different states of the system,
which usually is very large, $E_n$ is the energy of the state, and $\beta = 1/k_BT$ where $k_B$ is Boltzmann's constant and $T$ is the temperature.
Disregarding exceptional cases such as the two-dimensional Ising model, for a nontrivial many-body system the partition function $Z$ is unknown.
Hence, $p_n$ is not known. We can now pose the question "Can we construct a simulation algorithm that generates states according to the
unknown probability distribution $(p_1, \ldots , p_N)$?" As already mentioned, an affirmative answer to this question was given a long time ago by
Metropolis {\it et al.}~\cite{LAND00}. The basic idea is to construct a dynamical system, a Markov chain or master equation that samples
the space of $N$ states such that in the long run, the frequency with which this system visits the state $n$ approaches $p_n$ with probability one.

\subsection{The event-based simulation approach}
The basic ideas of the simulation approach that we review in the present paper are that
(i) we stick to what we know about the experiment, that is we consider the experimental configuration, its outcome and its data
analysis procedure as input for constructing the simulation algorithm;
(ii) we try to invent a set of simple rules that generates the same type of data as those recorded in an experiment
while reproducing the averages predicted by quantum theory;
(iii) we keep compatibility with macroscopic concepts.
Our event-based simulation approach is unconventional in that it does not require
knowledge of the wave amplitudes obtained by first solving a wave mechanical problem.
Evidently, as outlined below, our event-based approach requires a departure from the traditional way of describing physical phenomena,
namely in terms of locally causal, modular, adaptive, classical (non-Hamiltonian) dynamical systems.

Our event-based approach has successfully been used for discrete-event simulations of the single beam splitter and
Mach-Zehnder interferometer experiment
of Grangier {\sl et al.}~\cite{GRAN86} (see Refs.~\cite{RAED05d,RAED05b,MICH11a}),
Wheeler's delayed choice experiment of Jacques {\sl et al.}~\cite{JACQ07}
(see Refs.~\cite{ZHAO08b,MICH10a,MICH11a}),
the quantum eraser experiment of Schwindt {\sl et al.}~\cite{SCHW99} (see Ref.~\cite{JIN10c,MICH11a}),
two-beam single-photon interference experiments and the single-photon interference experiment with
a Fresnel biprism of Jacques {\sl et al.}~\cite{JACQ05} (see Ref.~\cite{JIN10b,MICH11a}),
quantum cryptography protocols (see Ref.~\cite{ZHAO08a}),
the Hanbury Brown-Twiss experiment of Agafonov {\sl et al.}~\cite{AGAF08} (see Ref.~\cite{JIN10a,MICH11a}),
universal quantum computation (see Ref.~\cite{RAED05c,MICH05}),
Einstein-Podolsky-Rosen-Bohm (EPRB)-type of experiments of Aspect {\sl et al.}~\cite{ASPE82a,ASPE82b}
and Weihs {\sl et al.}~\cite{WEIH98} (see Refs.~\cite{RAED06c,RAED07a,RAED07b,RAED07c,RAED07d,ZHAO08,MICH11a}),
and the propagation of electromagnetic plane waves through homogeneous thin films and stratified media (see Ref.~\cite{TRIE11,MICH11a}).
An extensive review of the simulation method and its applications is given in Ref.~\cite{MICH11a}.
Interactive demonstration programs, including source codes, are available for download~\cite{COMPPHYS,MZI08,DS08}.
A computer program to simulate single-photon EPRB experiments can be found in Ref.~\cite{RAED07a}.
So far the event-based simulation approach has not been used to simulate the phenomena of diffraction and evanescent waves.

Given the above list of successful simulations, clearly an affirmative answer has been given to the primary question posed in section 1.2.
A more detailed look to the algorithms that have been designed indicates that they are not unique.
Regarding the efficiency of the algorithms we can say that (i) they are more efficient than their experimental counterparts
because idealized models of the experimental equipment are considered and (ii) their efficiency is limited by the computational power of existing
digital computers. Although the event-based simulation approach can be used to simulate a universal quantum computer~\cite{RAED05c,MICH05},
the so-called ``quantum speed-up'' cannot be obtained. This by itself is no surprise because the quantum speed-up
is the result of a mathematical construct in which each unitary operation on the state of the quantum computer is counted as
one operation and in which preparation and read-out of the quantum computer are excluded. Whether or not this mathematical
construct is realized in Nature is an open question.

\subsection{Event-based modeling and detection efficiency}
Applications in quantum information have increased the interest in single-particle detectors.
High detection efficiencies are essential in for example quantum cryptography and some Bell test experiments.
Single-particle detectors are often complex devices with diverse properties.
In our event-based simulation approach we model the main characteristics of these devices by very simple rules.
So far, we have designed two types of detectors, simple particle counters and adaptive threshold devices (see section 2.1).
The adaptive threshold detector can be employed in the simulation of all single-photon experiments we have considered so far but is
absolutely essential in the simulation of for example the two-beam single photon experiment.
The efficiency, which is the ratio of detected to emitted particles, of our model detectors is measured in an experiment
with one single point source emitting single particles that is placed far away from the detector. According to this
definition, the simple particle counter has an
efficiency of 100\% and the adaptive threshold detector has an efficiency of nearly 100\%.
Hence, these detectors are highly idealized versions of real single photon detectors. No absorption effects, dead times, dark counts or other
effects causing particle miscounts are simulated.

In laboratory experiments, the single particle detection process is often quantified in terms of the ratio of detected to emitted particles,
the overall detection efficiency. Evidently, the efficiency of the detector plays an important role in this overall detection efficiency
but is not the only determining factor. Also the experimental configuration in which the detector is used plays an important role.
Therefore, the experimenter usually choses the best overall performing single-particle detector for her or his particular
experiment.
Also in the event-based approach the experimental configuration plays an important role in the overall detection efficiency.
Although the adaptive threshold detectors are ideal and have a detection efficiency of nearly 100\%, the overall detection efficiency can
be much less than 100\% depending on the experimental configuration. For example, using adaptive
threshold detectors in a Mach-Zehnder
interferometry experiment leads to an overall detection efficiency of nearly 100\%, while using the same detectors in a single-photon
two-beam experiment (see section 2.1) leads to an overall detection efficiency of about 15\%~\cite{MICH11a,JIN10b}.

Also the data processing procedure which is applied after the data has been collected may have an influence on the final detection efficiency.
For example, in the single-photon EPRB experiment of Weihs {\sl et al.} a post-selection procedure with a time-coincidence window is employed
to group photons, detected in two different stations, into pairs~\cite{WEIH98}.
Clearly, this procedure omits detected photons and therefore reduces the final ratio of detected to emitted photons.
This is also the case in the event-based simulation of this experiment (see section 2.2).
Although simple particle counters with a 100\% detection efficiency are used and thus all emitted photons are accounted for during the data
collection process, the final detection efficiency is less than 100\% because some detection events are omitted in the post-selection
data procedure using a time-coincidence window.

In conclusion, even if ideal detectors with a detection efficiency of 100\% would be commercially available, then the overall detection efficiency
in a single-particle experiment could still be much less than 100\% depending on (i) the experimental configuration in which the detectors
are employed and (ii) the data analysis procedure that is used after all data has been collected.

\subsection{Event-based modeling and the interpretations of quantum theory}
This paper is not about interpretations or extensions of quantum theory.
The fact that there exist simulation algorithms that reproduce the statistical results of quantum theory
has no direct implications for the foundations of quantum theory.
The average properties of the data may be in perfect agreement with quantum theory but
the algorithms that generate such data are outside of the scope of what quantum theory can describe.
Nevertheless, one could say that the event-based simulation approach is in line with the ensemble-statistical interpretation
of quantum theory but also goes beyond this interpretation since the method is able to give a logical cause-and-effect description
of how the ensemble is generated event by event. This makes metaphysical interpretations~\cite{DEUT11,VEDR11} superfluous, at least
for those experiments that can be simulated by the event-based approach.

Of course, providing a description that goes beyond the statistical properties comes at a price.
As it is well-known, ``non-contextuality'', literally meaning ``being independent of the (experimental) measurement arrangement''
is one of the properties that makes quantum (Maxwell's) theory so widely applicable~\cite{HOME97}.
To go beyond a statistical description unavoidably requires contextuality~\cite{HOME97}.
Therefore, the discrete-event approach provides a complete description of individual events but
non-contextuality is lost.
Note that as long as no measurement is performed the event-based description is non-contextual.
For example, in the simulation of the single-photon EPRB experiment (see section 2.2)
the two photons leaving the source have random but opposite polarization.
The photons have a well-defined predetermined (non-contextual) polarization which corresponds to a certain vector ${\mathbf S}$.
When an observer makes a measurement on one photon using a polarizing beam splitter and two detectors,
the observer gets a response +1 or -1 depending
on the angle of the polarizing beam splitter. Hence, depending on the angle of the polarizing beam splitter the observer measures +1 or -1 while
the polarization vector ${\mathbf S}$ of the photon has a well-defined predetermined length and orientation.
Thus, in this case, the contextuality (dependence on measurement arrangment) stems
from the fact that the observer can only obtain partial information about the vector ${\mathbf S}$ when making a measurement.

Finally, it should be noted that although
the discrete-event algorithm can be given an interpretation as a realistic cause-and-effect description that
is free of logical difficulties and reproduces the statistical results of quantum theory,
at present the lack of relevant data makes it impossible to decide whether or not such algorithms are realized by Nature.
Only new, dedicated experiments that probe more than just the statistical properties
can teach us more about this intriguing question. Proposals for such experiments are discussed in Refs.~\cite{JIN10b,MICH12a}.

\section{Illustrative examples}\label{ILL}
As a detailed discussion of an extensive set of discrete-event rules
cannot be fitted in this short review, we have opted to present a detailed
description of the algorithms for some fundamental experiments
in quantum physics, namely interference of two coherent beams of particles
and two Bell test experiments, an Einstein-Podolsky-Rosen-Bohm (EPRB) experiment with single photons~\cite{WEIH98} and a
Bell's inequality interference experiment with single neutrons~\cite{HASE03}.
Our motivation to select the two-beam experiment with single photons as an example is that this experiment
with minimal equipment shows interference
in its purest form, that is without diffraction being involved.
The experiment demonstrates that single particles coming from two coherent beams can gradually build up an interference pattern
when the particles arrive one by one at a detector screen.
We choose the Bell test experiments as another example because the quantum
theoretical description of this type of experiments involves entanglement.
The single-photon EPRB experiment demonstrates that the two photons of a pair, post-selected by employing a time-coincidence window, can
be in an entangled state.
The neutron interferometry experiment shows that it is possible to create correlations between the spatial and spin degree of freedom of
neutrons which, within quantum theory, cannot be described by a product state meaning that the spin- and phase-degree-of-freedom are entangled.
In this experiment the neutrons are counted with a detector having a very high efficiency ($\approx 99\%$), thereby not suffering
from the so-called detection loophole.

Furthermore, we have chosen to present the close-to-simplest algorithms
that reproduce the results of quantum theory,
hoping that this will help the reader grasp the basic ideas.
For an extensive review of an event-based model for quantum optics
experiments, see Ref.~\cite{MICH11a}.
\subsection{Two-beam interference}\label{TBI}

\begin{figure}[t]
\begin{center}
\includegraphics[width=7cm]{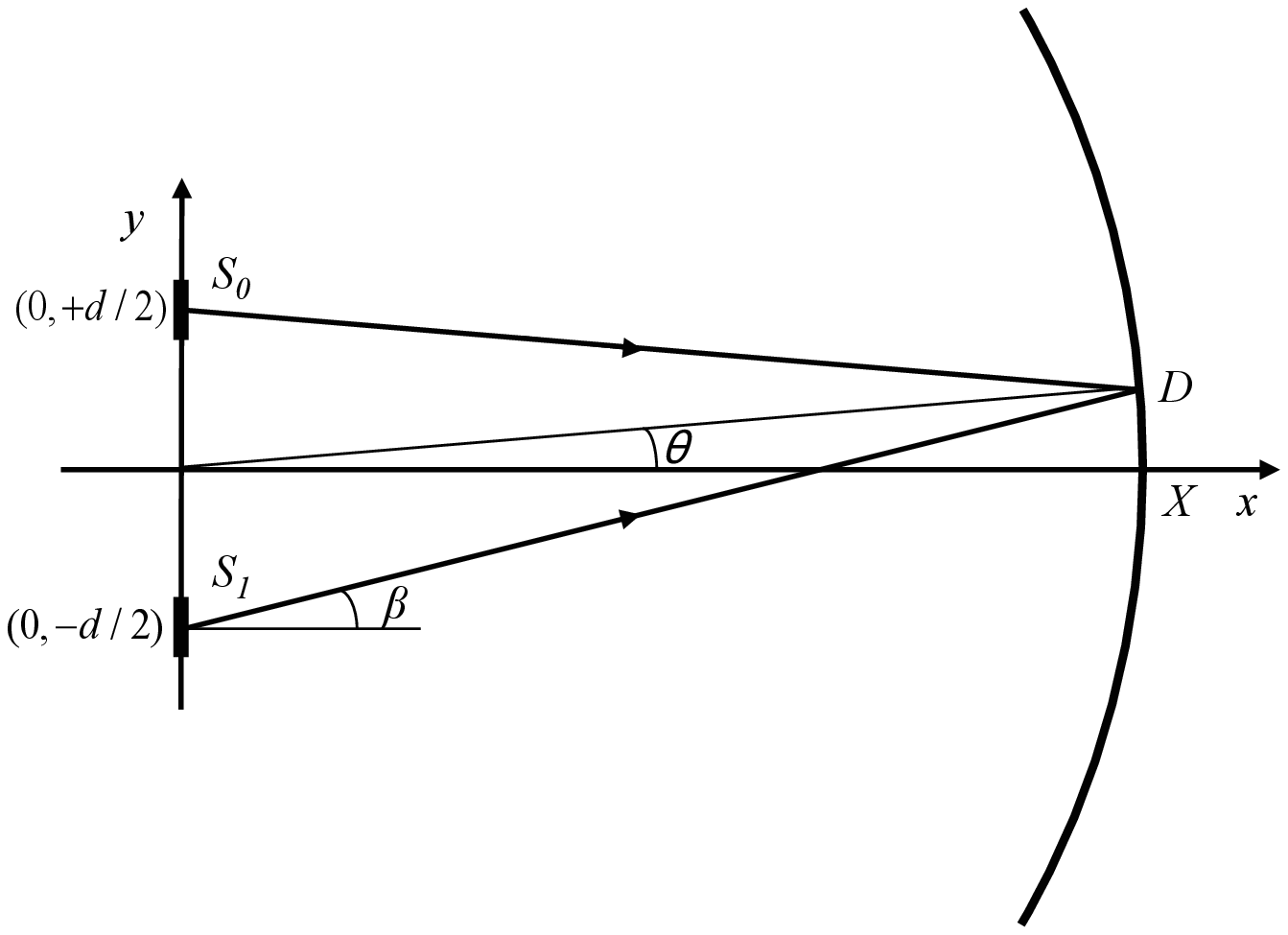}
\includegraphics[width=7cm]{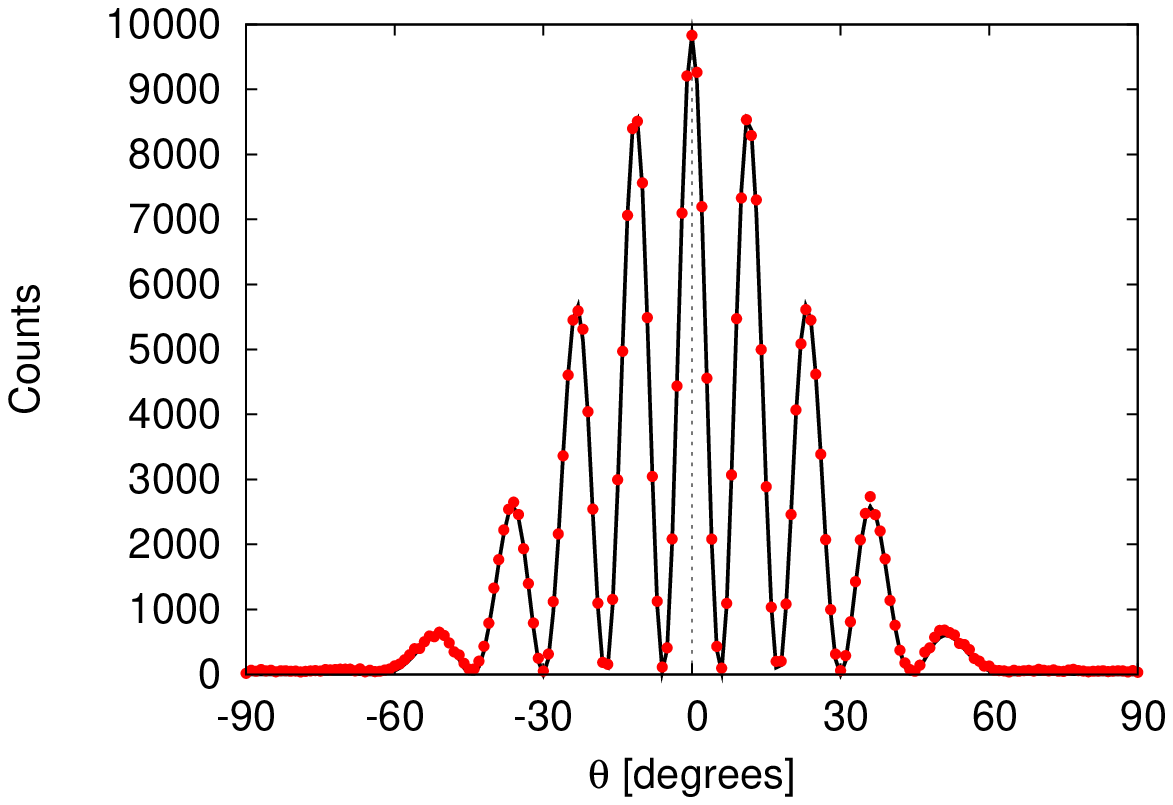}
\caption{%
Left: schematic diagram of a two-beam experiment with light sources $S_0$ and $S_1$ of width $a$,
separated by a center-to-center distance $d$.
These sources emit coherent, monochromatic light according to the light flux distribution
$
J(x,y)=\delta(x)\left[\Theta(a/2-|y-d/2|)+\Theta(a/2-|y+d/2|)\right]
$,
where $\Theta(.)$ denotes the unit step function
and with a uniform angular distribution, $\beta$ denoting the angle.
The light is recorded by detectors $D$ positioned on a semi-circle with radius $X$ and center $(0,0)$.
The angular position of a detector is denoted by $\theta$.
Right: detector counts (circles) as a function of $\theta$
as obtained from the \EBM\ simulation of the two-beam interference experiment depicted on the left.
The solid line is a least-square fit of the simulation data to the prediction of wave theory, Eq.~(\ref{tbi2}), with only one fitting parameter.
Simulation parameters: on average, each of the 181 detectors receives $10^4$ particles, $\gamma=0.99$, $a=c/f$, $d=5c/f$, $X=100c/f$, where $c$ denotes
the velocity and $f$ the frequency of the particles.
}
\label{exptbi}
\label{simtbi}
\end{center}
\end{figure}

In 1924, de Broglie introduced the idea that also matter, not just light, can exhibit wave-like properties~\cite{BROG25}.
This idea has been confirmed in various double-slit experiments with massive objects such as
electrons~\cite{JONS61,MERL76,TONO89,NOEL95}, neutrons~\cite{ZEIL88,RAUC00}, atoms~\cite{CARN91,KEIT91}
and molecules such as $C_{60}$ and $C_{70}$~\cite{ARND99,BREZ02}, all showing interference.
In some of the double-slit experiments~\cite{MERL76,TONO89,JACQ05}
the interference pattern is built up by recording individual clicks of the detectors.
According to Feynman, the observation that the interference patterns are built up event-by-event
is ``impossible, absolutely impossible to explain in any classical way
and has in it the heart of quantum mechanics''~\cite{FEYN65}.

Reading ``any classical way'' as ``any classical Hamiltonian mechanics way'',
Feynman's statement is difficult to dispute.
However, taking a broader view by allowing for dynamical systems that are outside the realm of classical
Hamiltonian dynamics, it becomes possible to model the gradual appearance of interference patterns
through a discrete-event simulation that does not make reference to concepts of quantum theory.
As a discrete-event simulation is nothing but a sequence of instructions
that changes the state of a macroscopic classical apparatus
(most conveniently a digital computer) in a prescribed manner,
the demonstration that we give in this section
provides a rational, logically consistent, common-sense explanation of
how the detection of individual objects that do not interact with each other
can give rise to the interference patterns that are being observed.
As there are several breeds of such models~\cite{JIN10b,MICH11a,WEBB11} that reproduce
the statistical distribution predicted by quantum theory,
only a new kind of experiment, specifically addressing this issue
can provide a verdict about the applicability of these models to the problem at hand~\cite{JIN10b}.

For the purpose of illustration, we consider the simple experiment sketched in Fig.~\ref{exptbi}(left).
This two-beam experiment can be viewed as a simplification of Young's double-slit experiment in which the slits
are regarded as the virtual sources $S_0$ and $S_1$~\cite{BORN64}.
In the two-beam experiment interference appears in its most pure form because in contrast to the
two-slit experiment the phenomenon of diffraction is absent.
The \EBM\ for this experiment has two types of events only: (i) the creation of one
particle at one of the sources and (ii) the detection of that particle by one of the detectors forming the screen.
We assume that all these detectors are identical and cannot communicate among each other
and do not allow for direct communication between the particles, implying
that this event-by-event model is locally causal by construction.
Then, if it is indeed true that individual particles build up the interference pattern one by one,
just looking at Fig.~\ref{exptbi}(left) leads to the logically unescapable
conclusion that the interference pattern can only be due to the internal operation of the detector~\cite{PFlE67}.
Detectors that simply count the incoming photons are not sufficient to explain the appearance of an interference pattern
and apart from the detectors there is nothing else that can cause the interference pattern to appear.
Making use of the statistical property of quantum theory one could assume that if a detector is replaced by another one as soon as it detects one photon,
one obtains a similar interference pattern if the detection events of all these different detectors are combined.
However, since there is no experimental evidence that confirms this assumption
and since our event-based approach is based on laboratory experimental setups and observations we do not consider this being a realistic option.
Thus, logic dictates that a minimal \EBM\ for the two-beam experiment requires an algorithm for
the detector that does a little more than just counting particles.

\subsubsection{Event-based model}
We now specify the model in sufficient detail such that the reader who is interested can reproduce
our results (a Mathematica implementation of a slightly more sophisticated algorithm~\cite{JIN10b}
can be downloaded from the Wolfram Demonstration Project web site~\cite{DS08}).

\begin{itemize}
\item{{\sl Source and particles:} %
The particles leave the source one by one, at positions $y$ drawn randomly from
a uniform distribution over the interval $[-d/2-a/2,-d/2+a/2]\cup[+d/2-a/2,+d/2+a/2]$.
The particle is regarded as a messenger, traveling in the direction given by the angle $\beta$, being a uniform pseudo-random number between $-\pi/2$ and $\pi/2$.
Each messenger carries with it a message
$\mathbf{e}(t)=(\cos 2\pi f t, \sin 2\pi f t)$
that is represented by a harmonic oscillator
which vibrates with frequency $f$ (representing the ``color'' of the light).
The internal oscillator is used as a clock to encode the time of flight $t$.
When a messenger is created, its time of flight is set to zero.
This pictorial model of a ``photon'' was used by Feynman to explain
quantum electrodynamics~\cite{FEYN85}.
The event-based approach goes one step further in that it specifies
in detail, in terms of a mechanical procedure, how the ``amplitudes''
that appear in the quantum formalism get added together.

The time of flight of the particles depends on the source-detector distance.
Here, we discuss as an example, the experimental setup with a semi-circular detection screen (see Fig.~\ref{exptbi}(left))
but in principle any other geometry for the detection screen can be considered.
The messenger leaving the source at $(0, y)$ under an angle $\beta$
will hit the detector screen of radius $X$ at a position determined by the angle $\theta$ given by
$\sin\theta=(y\cos^2\beta+\sin\beta\sqrt{X^2-y^2\cos^2\beta})/{X}$,
where $|y/X|< 1$. The time of flight is then given by
$t=\sqrt{X^2-2yX\sin\theta+y^2}/c$,
where $c$ is the velocity of the messenger.
The messages $\mathbf{e}(t)$ together with the explicit expression for the time of flight are the only input to the event-based algorithm.
}%
\item{{\sl Detector:} %
Microscopically, the detection of a particle involves very intricate dynamical processes~\cite{NIEU11b}.
In its simplest form, a light detector consists of a material that can be ionized by light.
This signal is then amplified, usually electronically, or in the case of a photographic plate
by chemical processes.
In Maxwell's theory, the interaction between the incident electric field ${\mathbf E}$ and
the material takes the form ${\mathbf P}\cdot{\mathbf E}$, where ${\mathbf P}$ is the polarization vector of the material~\cite{BORN64}.
Assuming a linear response, ${\mathbf P}(\omega)={\mathbf \chi}(\omega){\mathbf E}(\omega)$ for a monochromatic wave with
frequency $\omega$, it is clear that in the time domain, this relation expresses the fact that the material
retains some memory about the incident field, ${\mathbf \chi}(\omega)$ representing the memory kernel
that is characteristic for the material used.

In line with the idea that an event-based approach should use the simplest rules possible,
we reason as follows.
In the \EBM, the $k$th message ${\mathbf e}_k=(\cos 2\pi f t_k, \sin 2\pi f t_k)$ is taken
to represent the elementary unit of electric field ${\mathbf E}(t)$.
Likewise, the electric polarization ${\mathbf P}(t)$ of the material is represented by the
vector ${\mathbf p}_k = (p_{0,k},p_{1,k})$.
Upon receipt of the $k$th message this vector is updated according to the rule
\begin{equation}
	{\mathbf p}_k = \gamma {\mathbf p}_{k-1} + (1-\gamma) {\mathbf e}_k,
	\label{ruleofLM}
\end{equation}
where $0<\gamma<1$ and $k>0$.
Obviously, if $\gamma>0$, a message processor that operates according to the update rule Eq.~(\ref{ruleofLM}) has memory,
as required by Maxwell's theory.
It is not difficult to prove that as $\gamma\rightarrow1^-$, the internal vector ${\mathbf p}_{k}$
converges to the average of the time-series $\{{\mathbf e}_{1},{\mathbf e}_{2},\ldots\}$~\cite{JIN10b,MICH11a}.
The parameter $\gamma$ controls the precision
with which the machine defined by Eq.~(\ref{ruleofLM}) learns the average of the sequence of messages
${\mathbf e}_{1},{\mathbf e}_{2}, \ldots$
and also controls the pace at which new messages affect the internal state of the machine~\cite{RAED05d}.
Moreover, in the continuum limit (meaning many events per unit of time),
the rule Eq.~(\ref{ruleofLM}) translates into the constitutive equation of the Debye model of a dielectric~\cite{JIN10b},
a model used in many applications of Maxwell's theory~\cite{TAFL05}.
The learning process of the detector, governed by the rule Eq.~(\ref{ruleofLM}) can namely be viewed as a process that
proceeds in discrete time steps $\tau$, so that ${\mathbf p}_k={\mathbf p}(k\tau)={\mathbf p}(t)$ and
${\mathbf e}_k={\mathbf e}(k\tau)={\mathbf e}(t)$.
For small $\tau$, ${\mathbf p}_{k-1}={\mathbf p}(t)-\tau\partial {\mathbf p}(t)/\partial t +O(\tau^2)$
and hence
\begin{equation}
\frac{\partial {\mathbf p}(t)}{\partial t}\approx -\frac{1-\gamma}{\tau\gamma}{\mathbf p}(t) + \frac{1-\gamma}{\tau\gamma}{\mathbf e}(t).
\end{equation}
Letting the time step $\tau$, that is the time between the arrival of successive messages, approach zero ($\tau\rightarrow 0$),
letting $\gamma$ approach one ($\gamma\rightarrow 1^{-}$) and demanding that the resulting continuum equation
makes sense, leads to the following relation between $\tau$ and $\gamma$,
\begin{equation}
\lim_{\tau\rightarrow 0}\lim_{\gamma\rightarrow 1^-}\frac{1-\gamma}{\tau\gamma}=\Gamma,
\label{relax}
\end{equation}
with $0<\Gamma <\infty$.
Hence, $\partial {\mathbf p}(t)/\partial t=-\Gamma {\mathbf p}(t)+\Gamma {\mathbf e}(t)$ or in Fourier space
${\mathbf P}(\omega)=\chi(\omega){\mathbf E}(\omega)$, being the constitutive equation of the Debye model of a dielectric,
with Eq.~(\ref{relax}) giving an explicit expression for the relaxation time $1/\Gamma$ in terms of the parameters
of the event-based model.

After updating the vector ${\mathbf p}_k$, the processor uses the information stored in ${\mathbf p}_k$
to decide whether or not to generate a click.
As a highly simplified model for the bistable character of the real photodetector or
photographic plate, we let the machine generate a binary output signal $S_k$ according to
\begin{equation}
	S_k = \Theta({\mathbf p}^2_k-r_k),
	\label{thresholdofLM}
\end{equation}
where $\Theta(.)$ is the unit step function and $0\leq r_k <1$ is a uniform pseudo-random number.
Note that the use of random numbers is convenient but not essential~\cite{MICH11a}.
Since in experiment it cannot be known whether a photon has gone undetected, we discard the information about
the $S_k=0$ detection events and define the total detector count as
$N=\sum^{k}_{j=1}S_j$,
%
where $k$ is the number of messages received. $N$ is the number of clicks (one's) generated by the processor.

The efficiency of the detector model is determined by simulating an experiment
that measures the detector efficiency, which for a single-photon detector is defined
as the overall probability of registering a count if a photon arrives at the detector~\cite{HADF09}.
In such an experiment a point source emitting single particles is placed far away from a single detector.
As all particles that reach the detector have the same time of flight (to a good approximation), all the
particles that arrive at the detector will carry the same message which is encoding the time of flight.
As a result ${\mathbf p}_k$ (see Eq.~(\ref{ruleofLM})) rapidly converges to the vector corresponding to
this identical message, so that the detector clicks every time a photon arrives.
Thus, the detection efficiency, as defined for real detectors~\cite{HADF09}, for our detector model is very close
to 100\%.
Hence, the model is a highly simplified and idealized version of a single-photon detector.
However, although the detection efficiency of the detector itself may be very close to 100\%,
the overall detection efficiency, which is
the ratio of detected to emitted photons in the simulation of an experiment,
can be much less than one. This ratio depends on the
experimental setup.
}
\item{{\sl Simulation procedure:} %
Each of the detectors of the circular screen has a predefined spatial window within which it accepts messages.
As a messenger hits a detector, this detector updates its internal state ${\mathbf p}$,
(the internal states of all other detectors do not change)
using the message ${\mathbf e}_{k}$ and then generates the event $S_k$.
In the case $S_k=1$, the total count of the particular detector that was hit by the $k$th
messenger is incremented by one and the messenger itself is destroyed.
Only after the messenger has been destroyed, the source is allowed to send a new messenger.
This rule ensures that the whole simulation complies with Einstein's criterion of local causality.
This process of creating and destroying messengers is repeated many times, building up the
interference pattern event by event.
}
\end{itemize}

\subsubsection{Simulation results}
In Fig.~\ref{simtbi}(right), we present simulation results for a representative case for which the analytical solution from wave theory is known.
Namely, in the Fraunhofer regime ($d\ll X$), the analytical expression for the light intensity at the detector on a circular screen with radius $X$ is given by~\cite{BORN64}
\begin{equation}
I(\theta) = A\left.{\sin^2\frac{qa\sin\theta}{2}}\cos^2\frac{qd\sin\theta}{2}\right/\left( \frac{qa\sin\theta}{2}\right)^2
,
\label{tbi2}
\end{equation}
where $A$ is a constant, $q=2\pi f/c$ denotes the wavenumber with $f$ and $c$ being the frequency and velocity of the light, respectively,
and $\theta$ denotes the angular position of the detector $D$ on the circular screen, see Fig.~\ref{exptbi}(left).
Note that Eq.~(\ref{tbi2}) is only used for comparison with the simulation data and is by no means input to the model.
From Fig.~\ref{simtbi}(right) it is clear that the \EBM\ reproduces the results of wave theory
and this without taking recourse of the solution of a wave equation.

As the detection efficiency of the event-based detector model is very close to 100\%,
the interference patterns generated by the \EBM\ cannot be attributed
to inefficient detectors.
It is therefore of interest to take a look at the ratio of detected to emitted photons, the overall detection efficiency, and
compare the detection counts, observed in the \EBM\ simulation
of the two-beam interference experiment, with those observed in a real experiment with single photons~\cite{JACQ05}.
In the simulation that yields the results of Fig.~\ref{simtbi}, each of the 181 detectors making up the
detection area is hit on average by ten thousand photons and the total number of clicks generated
by the detectors is 296444. Hence, the ratio of the total number of detected to emitted photons is
of the order of 0.16, two orders of magnitude larger than
the ratio $0.5\times 10^{-3}$ observed in single-photon interference experiments~\cite{JACQ05}.

\subsubsection{What is the working principle?}
In our event-based approach the simple particle counters and the adaptive threshold detectors
are ideal detectors with a detection efficiency of (nearly) 100\%.
In the case of the two-beam interference experiment considered in section 2.1,
using simple particle counters would not result in an interference pattern.
These detectors simply produce a click for each incoming
photon and do nothing with the information encoded in the messages ${\mathbf e}$ carried by the particles.
These messages contain information about the time of flight of the particles, that is about the distance travelled
by the particles from one of the two sources to one of the detectors constituting the circular detection screen.
It is precisely the difference in the times of flight (or the phase differences) which is important in the
generation of an interference pattern. Since, in the single-photon two-beam experiment the detectors are the only apparatuses available
in the experiment that can process this information (there are no other apparatuses present except for the source)
we necessarily need to employ an algorithm for the detector that exploits this information in order to produce the clicks
that gradually build up the interference pattern. A collection of about two hundred independent adaptive threshold detectors defined by
Eq.~(\ref{ruleofLM}) and Eq.~(\ref{thresholdofLM}) and each with a
detection efficiency of nearly 100\% is capable of doing this.
As pointed out earlier, the reason why, in this particular experiment, this is possible is that not every particle that
impinges on the detector yields a click.

Note that to simulate the interference pattern observed in single-photon Mach-Zehnder interferometry experiments
it is possible, but by no means necessary, to use the adaptive threshold detectors which do not necessarily produce a click for each incoming photon~\cite{MICH11a}.
Indeed, it suffices to use simple particle counters which produce a click for each incoming photon~\cite{RAED05d,RAED05b}.
In the simulation of the Mach-Zehnder interferometry experiment the beam splitters process the information about the time of flight of the particles~\cite{RAED05d,RAED05b,MICH11a}.
Whether or not the detectors also process this information has no influence on the generation of the interference pattern~\cite{RAED05d,RAED05b,MICH11a}.
Hence, in this case the number of detected photons is equal to the number of emitted photons in the simulation.

\subsection{Einstein-Podolsky-Rosen-Bohm (EPRB) experiment with single photons}\label{model}\label{simulationmodel}

The EPRB experiment with photons, carried out by Weihs {\it et al.}~\cite{WEIH98,WEIH00},
is taken as a concrete example to illustrate how to construct an \EBM\ that
reproduces the predictions of \QT\ for the single and two-particle averages
for a quantum system of two spin-1/2 particles in the singlet state and
a product state~\cite{ZHAO08,MICH11a}, without making reference to concepts of quantum theory.
Recall that the quantum theoretical descriptions of the EPRB experiment with photons
or with spin-1/2 particles are identical.

In short, the experiment goes as follows.
A source emits pairs of photons.
Each photon of a pair travels to an observation station in which it is manipulated and detected.
The two stations are assumed to be identical.
They are separated spatially and temporally, preventing the observation at station 1 (2) to have a causal effect on the
data registered at station $2$ (1)~\cite{WEIH98}.
As the photon arrives at station $i=1,2$, it passes through an electro-optic modulator (EOM)
which rotates the polarization of the photon by an angle depending on the voltage applied to the modulator.
These voltages are controlled by two independent binary random number generators.
A polarizing beam splitter sends the photon to one of the two detectors.
The station's clock assigns a time-tag to each generated signal.

The firing of a detector is regarded as an event.
At the $n$th event, the data recorded on a hard disk at station $i=1,2$
consists of $x_{n,i}=\pm 1$, specifying which of the two detectors fired,
the time tag $t_{n,i}$ indicating the time at which a detector fired,
and the two-dimensional unit vector $\alpha_{n,i}$ that represents the rotation
of the polarization by the EOM at the time of detection.
Hence, the set of data collected at station $i=1,2$ may be written as
\begin{eqnarray}
\Upsilon_i=\left\{ {x_{n,i} =\pm 1,t_{n,i},\alpha_{n,i} \vert n =1,\ldots ,N_i } \right\}
\label{eprb1}
.
\end{eqnarray}
In the experiment, the data $\{\Upsilon_1,\Upsilon_2\}$ is analyzed
after the data of a particular run has been collected~\cite{WEIH98}.
Adopting the procedure employed by Weihs {\sl et al.}~\cite{WEIH98,WEIH00},
we identify coincidences by comparing the time differences
$t_{n,1}-t_{m,2}$ with a window $W$ where $n=1,\ldots,N_1$ and $m=1,\ldots,N_2$.
By definition, for each pair of rotation angles $a$ and $b$ of the EOMs,
the number of coincidences between detectors $D_{x,1}$ ($x =\pm $1) at station 1 and
detectors $D_{y,2}$ ($y =\pm $1) at station 2 is given by
\begin{eqnarray}
C_{xy}=C_{xy}(a,b)&=&
\sum_{n=1}^{N_1}
\sum_{m=1}^{N_2}
\delta_{x,x_{n ,1}} \delta_{y,x_{m ,2}}
\delta_{a ,\alpha_{n,1}}\delta_{b,\alpha_{m,2}}
\Theta(W-\vert t_{n,1} -t_{m ,2}\vert)
,
\label{Cxy}
\end{eqnarray}
where $\Theta(.)$ denotes the unit step function.
In Eq.~(\ref{Cxy}) the sum over all events has to
be carried out such that each event (= one detected photon) contributes only once.
Clearly, this constraint introduces some ambiguity in the counting procedure as there
is a priori, no clear-cut criterion to decide which events at stations $i=1$ and $i=2$ should
be paired. One obvious criterion might be to choose the pairs such that $C_{xy}$ is maximum~\cite{WEIH00}
but, such a criterion renders the data analysis procedure (not the
data production) acausal. It is trivial though to analyse the data generated by
the experiment of Weihs {\sl et al.} such that conclusions do not suffer from this artifact~\cite{RAED12}.
The correlation of the two dichotomic variables $x$ and $y$ is defined as
\begin{eqnarray}
E(a,b)&=&\frac{C_{++}+C_{--}-C_{+-}-C_{-+}}{C_{++}+C_{--}+C_{+-}+C_{-+}}
,
\label{eprb5b}
\end{eqnarray}
where the denominator is the sum of all coincidences.
In general, the two-particle averages $E(a,b)$, and the total number of the coincidences
not only depend on the directions $a$ and $b$ but also on
the time window $W$ used to identify the coincidences.
For later use it is expedient to introduce the function~\cite{CLAU69}

\begin{equation}
S\equiv S(a,b,a^{\prime},b^{\prime})=
E(a,b)-E(a,b^{\prime})+E(a^{\prime},b)+E(a^{\prime},b^{\prime}).
\label{funcS}
\end{equation}

Local-realistic treatments of the EPRB experiment usually assume that the expression for the correlation,
as measured in the experiment, is given by~\cite{BELL93}
\begin{eqnarray}
\label{CxyBell}
C_{xy}^{(\infty)}(a,b)&=&\sum_{n=1}^N\delta_{x,x_{n ,1}} \delta_{y,x_{n ,2}}
\delta_{a ,\alpha_{n,1}}\delta_{b,\alpha_{n,2}}
,
\end{eqnarray}
which is obtained from Eq.~(\ref{Cxy}) (in which each photon contributes only once) by
assuming that $N=N_1=N_2$, pairs are defined by $n=m$, and by
taking the limit $W\rightarrow\infty$.
However, the working hypothesis that
the value of $W$ should not matter because the time window only serves to identify pairs
may not apply to real experiments.
The analysis of the data of the experiment of Weihs {\sl et al.} shows that
the average time between pairs of photons is of the order of $30\mus$ or more,
much larger than the typical values (of the order of a few nanoseconds)
of the time-window $W$ used in the experiments~\cite{WEIH00}.
In other words, in practice, the identification of photon pairs
does not require the use of $W$'s of the order of a few nanoseconds.
The small value of $W$ is required to maximize $|S|$ and is not really required for the data
to violate Bell's inequality $|S|\le 2$.
In other words, depending on the value of $W$, chosen by the experimenter when analyzing the data,
the inequality $|S|\le 2$ may or may not be violated.
Hence, also the conclusion about the state of the system depends on the value of $W$, which turns $W$ into a so-called
context parameter.
Analysis of the data of the experiment by Weihs {\it et al.} shows that $W$ can be as large as 150 ns for the Bell inequality to be violated~\cite{RAED12}
and in the time-stamping EPRB experiment of Ag\"uero {\it et al.}~\cite{AGUE09} $|S|\le 2$ is clearly violated for $W<9\mus$.
Hence, the use of a time-coincidence window does not create a ``loophole''.
Nevertheless, very often it is mentioned that these single-photon Bell test experiments suffer from the fair sampling loophole,
being the result of the usage of a time window $W$ to filter out coincident photons or being the result of the usage of inefficient detectors~\cite{ADEN07}.
The detection loophole was first closed in an experiment with two entangled trapped ions~\cite{Rowe01} and later in a single-neutron
interferometry experiment~\cite{HASE03} and in an experiment with two entangled qubits~\cite{ANSM10}.
However, the latter three experiments are not Bell test experiments performed according
to the CHSH protocol~\cite{CLAU69} because the two degrees of freedom
are not manipulated and measured independently.

The narrow time window $W$ in the experiment by Weihs {\it et al.} mainly acts as a filter that selects pairs
of which the individual photons differ in their time tags by the order of nanoseconds.
The possibility that such a filtering mechanism can lead to correlations
that are often thought to be a characteristic of quantum systems only was,
to our knowledge, first pointed out by P. Pearle~\cite{PEAR70} and later by A. Fine~\cite{FINE82},
opening the route to a description in terms of locally causal, classical models.
A concrete model of this kind was proposed by S. Pascazio who showed
that his model approximately reproduces the correlation of the singlet state~\cite{PASC86}
with an accuracy that seems beyond what is experimentally achievable to date.
Larson and Gill showed that Bell-like inequalities need to be modified
in the case that the coincidences are determined by a time-window filter~\cite{LARS04},
and models that exactly reproduce the results of quantum theory for the singlet
and uncorrelated state were found~\cite{RAED06c,RAED07b,ZHAO08,MICH11a}.
Here, we closely follow Refs.~\cite{RAED07b,ZHAO08}.

\subsubsection{Event-based model}
A minimal, discrete-event simulation model of the EPRB experiment by Weihs {\sl et al.} 
requires a specification of the information carried by the particles,
of the algorithm that simulates the source and
the observation stations, and of the procedure to analyze the data.

\begin{itemize}
\item{{\sl Source and particles:}
Each time, the source emits two particles that carry a vector
${\bf S}_{n,i}=(\cos(\xi_{n}+(i-1)\pi/2) ,\sin(\xi_{n}+(i-1)\pi/2))$,
representing the polarization of the photons.
This polarization is completely characterized by the angle $\xi _{n}$ and
the direction $i=1,2$ to which the particle moves.
A uniform pseudo-random number generator is used to pick the angle $0\le\xi _{n}<2\pi$.
Clearly, the source emits two particles with a mutually orthogonal, hence correlated but otherwise
random polarization.
Note that for the simulation of this experiment it is not necessary that the particles carry information about the phase $2\pi ft_{i,n}$.
In this case the time of flight $t_{i,n}$ is determined by the time-tag model (see below).
}
\item{{\sl Electro-optic modulator:}
The EOM in station $i=1,2$ rotates the polarization of the incoming particle by an angle $\alpha_i$,
that is its polarization angle becomes $\xi^{\prime}_{n,i}\equiv\mathrm{EOM}_i(\xi_{n}+(i-1)\pi/2,\alpha_i)=\xi_{n}+(i-1)\pi/2-\alpha_i$ symbolically.
Mimicking the experiment of Weihs {\sl et al.} in which $\alpha_1$ can take the values $a,a'$ and $\alpha_2$ can take the values $b,b'$,
we generate two binary uniform pseudo-random numbers $A_i=0,1$ and use them
to choose the value of the angles $\alpha_i$, that is
$\alpha_1=a(1-A_1)+a'A_1$ and $\alpha_2=b(1-A_2)+b'A_2$.
}
\item{{\sl Polarizing beam splitter:}
The simulation model for a polarizing beam splitter is defined by the rule
\begin{eqnarray}
x_{n,i}=\left\{
\begin{array}{lll}
+1 & \mbox{if} & r_n\le \cos^2\xi^{\prime}_{n,i}\\
-1 & \mbox{if} & r_n > \cos^2\xi^{\prime}_{n,i}
\end{array}
\right.
,
\label{sg1}
\end{eqnarray}
where $0< r_n<1$ are uniform pseudo-random numbers.
It is easy to see that for fixed $\xi^{\prime}_{n,i}=\xi^{\prime}_i$, this rule generates
events such that
%
the distribution of events complies with Malus law.
}
\item{{\sl Time-tag model:} %
As is well-known, as light passes through an EOM (which is essentially a tuneable wave plate), it experiences a retardation
depending on its initial polarization and the rotation by the EOM.
However, to our knowledge, time delays caused by retardation properties of waveplates, being components of various
optical apparatuses, have not yet been explicitly measured for single photons.
Therefore, in the case of single-particle experiments, we hypothesize that for each particle this delay is
represented by the time tag~\cite{RAED07b,ZHAO08}
\begin{eqnarray}
t_{n,i}&=&\lambda(\xi^{\prime}_{n,i}) r^{\prime}_n
,
\label{sg3b}
\end{eqnarray}
that is, the time tag is distributed uniformly ($0<r^{\prime}_n <1$ is a uniform pseudo-random number) over the interval $[0, \lambda(\xi^{\prime}_{n,i})]$.
For $\lambda(\xi^{\prime}_{n,i})=T_0\sin^4 2\xi^{\prime}_{n,i}$ this time-tag model, in combination with the model
of the polarizing beam splitter, rigorously reproduces the results
of quantum theory of the EPRB experiments in the limit $W\rightarrow0$~\cite{RAED07b,ZHAO08}.
We therefore adopt the expression $\lambda(\xi^{\prime}_{n,i})=T_0 \sin^4 2\xi^{\prime}_{n,i}$
leaving only $T_0$ as an adjustable parameter.
}
\item{{\sl Detector:} %
The detectors are ideal particle counters, meaning that they produce a click for each incoming particle.
Hence, we assume that the detectors have 100\% detection efficiency.
Note that also adaptive threshold detectors can be used (see section 1.4) equally well~\cite{MICH11a}.
}
\item{{\sl Simulation procedure:} %
The simulation algorithm generates the data sets $\Upsilon_i$, similar to the ones obtained in the experiment
(see Eq.~(\ref{eprb1})).
In the simulation, it is easy to generate the events such that $N_1=N_2$.
We analyze these data sets in exactly the same manner as the experimental data are analyzed, implying that we
include the post-selection procedure to select photon pairs by a time-coincidence window $W$.
The latter is crucial for our simulation method to give results that are very similar to those observed in a laboratory experiment.
Although in the simulation the ratio of detected to emitted photons is equal to one, the final detection efficiency is reduced
due to the time-coincidence post-selection procedure.
}
\end{itemize}

\begin{figure}[t]
\centering
\includegraphics[width=7cm]{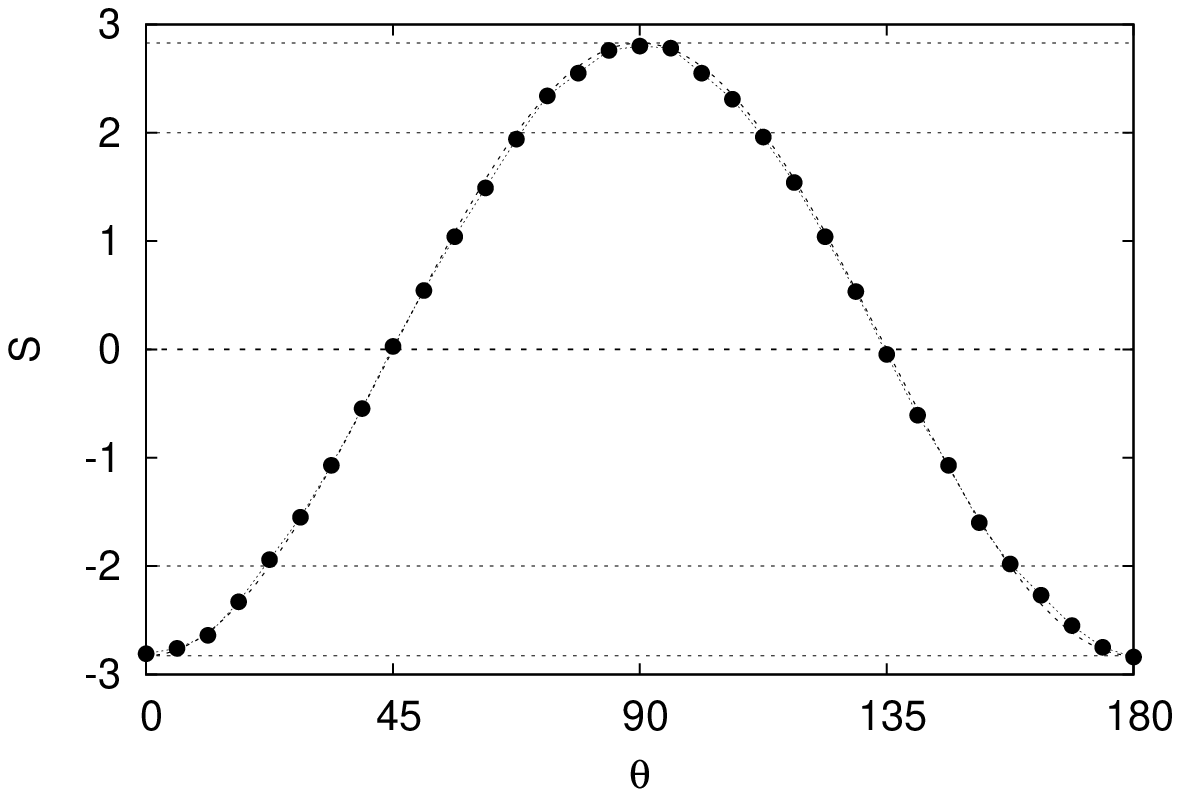}
\includegraphics[width=7cm]{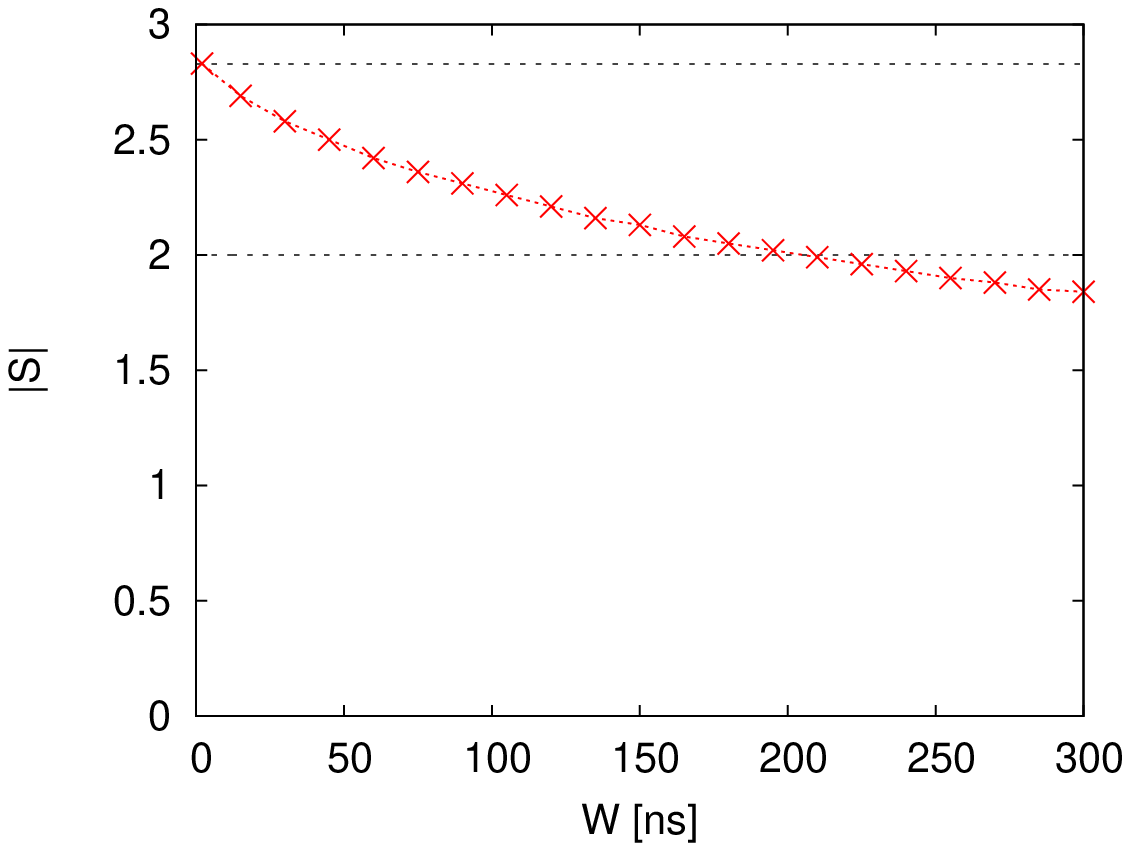}
\caption{
Simulation results produced by the locally causal, discrete-event simulation
model using the time-tag model Eq.~(\ref{sg3b}).
Left: $S$ as a function of $\theta$ with
$a=\theta$, $a'=\pi/4+\theta$, $b=\pi/8$ and $b'=3\pi/8$ for a time window $W=2\ns$.
The solid line connecting the circles is the result $-2\sqrt{2}\cos2\theta$ predicted by \QT.
The dashed lines represent the maximum value for a quantum system of two spin-1/2 particles
in a separable (product) state ($|S|=2$) and in a singlet state ($|S|=2\sqrt{2}$), respectively.
Right:
$|S|$ as a function of $W$ for
$a=0$, $a'=\pi/4$, $b=\pi/8$ and $b'=3\pi/8$.
For $W\approx 200\ns$, the system changes from ``quantum-like'' ($|S|>2$)
to ``classical like'' ($|S|\le2$).
Simulation parameters: for each value of $\theta$, the number of pairs generated is $3\times10^5$ (roughly the same
as in experiment~\cite{WEIH98}) and $T_0=2000\ns$, the adjustable parameter in the time-tag model Eq.~(\ref{sg3b}).
}
\label{fig.S0}
\end{figure}

This algorithm fully complies with
Einstein's criterion of local causality on the ontological level. Once the particles
leave the source, an action at observation station 1 (2) can, in no way,
have a causal effect on the outcome of the measurement at observation
station 2 (1).

\subsubsection{Simulation results}
In Fig.~\ref{fig.S0} we present some typical simulation results
for the function $S$,
showing that the \EBM\ reproduces the
predictions of \QT\ for the singlet state.
The single-particle averages $E_1(a)$ and $E_2(b)$ (data not shown) are zero up to
the usual statistical fluctuations and do not show any statistically
relevant dependence on $b$ or $a$, respectively.
Additional results, mimicking quantum systems in the singlet
and product state, as well as a rigorous probabilistic treatment of the model
can be found in~\cite{ZHAO08}.
For $W=50\ns$ (data not shown), we find $|S|=2.62$ which compares very well with the values between 2 and 2.57
extracted from different data sets produced by the experiment of Weihs {\sl et al.}
In all cases, the distribution of time-tag differences (data not shown)
is sharply peaked and displays long tails, in qualitative agreement with experiment~\cite{WEIH00}.

From Fig.~\ref{fig.S0}(right), it follows that a violation of the Bell inequality $|S|\le2$
depends on the choice of $W$, a parameter which is absent in the
quantum theoretical description of the EPRB thought experiment.
There are two limiting cases for which $S$ become independent of $W$.
If $W\rightarrow\infty$, it is impossible to let a digital computer violate the inequality $|S|\le2$
without abandoning the rules of Boolean logic or arithmetic~\cite{RAED11a}.
For relatively small $W$ ($W<150$ns), the inequality $|S|\le2$ may be violated.
When $W\rightarrow0$ the discrete-event models which generate the same type
of data as real EPRB experiments,
reproduce exactly the single- and two-spin averages of the singlet state and therefore also violate the
inequality $|S|\le2$.
Obviously, as the discrete-event model does not rely on any concept of quantum theory, a violation of the inequality $|S|\le2$ does not
say anything about the ``quantumness'' of the system under observation~\cite{KARL09,KARL10,RAED11a}.
Similarly, a violation of this inequality cannot say anything about locality and realism~\cite{KARL09,KARL10,RAED11a,NIEU11}.
Clearly, the \EBM\ is contextual.
The fact that the \EBM\ reproduces, for instance, the correlations of the singlet state
without violating Einstein's local causality criterion suggests that the
data $\{x_{n,1},x_{n,2}\}$ generated by the \EBM\ cannot be represented by a single Kolmogorov probability space.
This complies with the idea that contextual, non-Kolmogorov models can lead to
violations of Bell's inequality without appealing to nonlocality or nonobjectivism~\cite{KHRE09,KHRE11}.

\subsubsection{Why is Bell's inequality violated?}
In Ref.~\cite{ZHAO08}, we have presented a probabilistic description of our simulation model that (i) rigorously proves that
for up to first order in $W$ it exactly reproduces the single particle averages and the two-particle correlations of
quantum theory for the system under consideration; (ii) illustrates how the presence of the time-window $W$ introduces
correlations that cannot be described by the original Bell-like ``hidden-variable'' models~\cite{BELL93}.

The time-coincidence post-selection procedure with the time-window $W$ filters out the ``coincident'' photons based on the time-tags
$t_{n,i}$ thereby reducing the final detection efficiency to less than 100\%, although in the simulation a measurement
always returns a +1 or -1 for both photons in a pair (100\% detection efficiency of the detectors).
Hence, even in case of a perfect detection process the data set that is finally retained consists only of a subset of the entire
ensemble of correlated photons that was emitted by the source.
In other words, the use of a time-coincidence window destroys the ``fair-sampling hypothesis''.

We briefly elaborate on point (ii) (see Ref.~\cite{ZHAO08} for a more extensive discussion).
Let us assume that there exists a probability $P(x_1,x_2,t_1,t_2|\alpha_1,\alpha_2)$ to observe the data $\{x_1,x_2,t_1,t_2\}$
conditional on $\{\alpha_1,\alpha_2\}$.
The probability $P(x_1,x_2,t_1,t_2|\alpha_1,\alpha_2)$ can always be expressed as an integral over the mutually exclusive events
$\xi_1$, $\xi_2$, representing the polarization of the photons

\begin{equation}
P(x_1,x_2,t_1,t_2|\alpha_1,\alpha_2)=\frac{1}{4\pi^2}\int_0^{2\pi}\int_0^{2\pi}P(x_1,x_2,t_1,t_2|\alpha_1,\alpha_2,\xi_1,\xi_2)
P(\xi_1,\xi_2|\alpha_1,\alpha_2)d\xi_1d\xi_2.
\label{propmod1}
\end{equation}
According to Eq.~(\ref{sg1}) and Eq.~(\ref{sg3b}),
in the probabilistic version of our simulation model, for each event, (i) the values of
$x_1$, $x_2$, $t_1$, and $t_2$ are mutually
independent random variables, (ii) the values of $x_1$ and $t_1$ ($x_2$ and $t_2$)
are independent of $\alpha_2$ and $\xi_2$ ($\alpha_1$ and $\xi_1$),
(iii) $\xi_1$ and $\xi_2$ are independent of $\alpha_1$ or $\alpha_2$.
With these assumptions Eq.~(\ref{propmod1}) becomes

\begin{eqnarray}
P(x_1,x_2,t_1,t_2|\alpha_1,\alpha_2)
&\stackrel{\rm{(i)}}{=}&\frac{1}{4\pi^2}\int_0^{2\pi}\int_0^{2\pi}P(x_1,t_1|\alpha_1,\alpha_2,\xi_1,\xi_2)P(x_2,t_2|\alpha_1,\alpha_2,\xi_1,\xi_2)\nonumber\\
&&\times P(\xi_1,\xi_2|\alpha_1,\alpha_2)d\xi_1d\xi_2\nonumber\\
&\stackrel{\rm{(ii)}}{=}&\frac{1}{4\pi^2}\int_0^{2\pi}\int_0^{2\pi}P(x_1,t_1|\alpha_1,\xi_1)P(x_2,t_2|\alpha_2,\xi_2)\nonumber\\
&&\times P(\xi_1,\xi_2|\alpha_1,\alpha_2)d\xi_1d\xi_2\nonumber\\
&\stackrel{\rm{(i)}}{=}&\frac{1}{4\pi^2}\int_0^{2\pi}\int_0^{2\pi}P(x_1|\alpha_1,\xi_1)P(t_1|\alpha_1,\xi_1)P(x_2|\alpha_2,\xi_2)P(t_2|\alpha_2,\xi_2)\nonumber\\
&&\times P(\xi_1,\xi_2|\alpha_1,\alpha_2)d\xi_1d\xi_2\nonumber\\
&\stackrel{\rm{(iii)}}{=}&\frac{1}{4\pi^2}\int_0^{2\pi}\int_0^{2\pi}P(x_1|\alpha_1,\xi_1)P(t_1|\alpha_1,\xi_1)
P(x_2|\alpha_2,\xi_2)\nonumber\\
&&\times P(t_2|\alpha_2,\xi_2)P(\xi_1,\xi_2)d\xi_1d\xi_2,
\label{propmod2}
\end{eqnarray}
which is the probabilistic description of our simulation model.
The probabilistic model Eq.~(\ref{propmod2}) is identical to the local-realist model
used by Larsson and Gill in their derivation of a CHSH inequality with time-coincidence restriction~\cite{LARS04}.

According to our simulation model, the probability distributions that describe the polarizers are given by
$P(x_i|\alpha_i,\xi_i)=[1+x_i\cos 2 (\alpha_i -\xi_i)]/2$ and
those for the time-delays $t_i$ that are distributed randomly over the interval $[0,\lambda (\xi_i +(i-1)\pi/2 -\alpha_i)]$
are given by
$P(t_i|\alpha_i,\xi_i)=\theta (t_i)
\theta (\lambda (\xi_i +(i-1)\pi/2 -\alpha_i)-t_i)/\lambda (\xi_i +(i-1)\pi/2 -\alpha_i)$.
In the experiment~\cite{WEIH98} and therefore also in our simulation model, the events are selected using a time window $W$ that
the experimenters try to make as small as possible~\cite{WEIH00}.
Accounting for the time window, that is multiplying Eq.~(\ref{propmod2}) by a step function and integrating over all $t_1$ and $t_2$,
the expression for the probability for observing the event $(x_1,x_2)$ reads

\begin{equation}
P(x_1,x_2|\alpha_1,\alpha_2)=\int_0^{2\pi}\int_0^{2\pi}P(x_1|\alpha_1,\xi_1)P(x_2|\alpha_2,\xi_2)
\rho(\xi_1,\xi_2|\alpha_1,\alpha_2)d\xi_1d\xi_2,
\label{propmod3}
\end{equation}
where the probability density $\rho(\xi_1,\xi_2|\alpha_1,\alpha_2)$ is given by
\begin{eqnarray}
&&\rho(\xi_1,\xi_2|\alpha_1,\alpha_2)=
\nonumber\\
&&\frac{\int_{-\infty}^{+\infty}\int_{-\infty}^{+\infty}P(t_1|\alpha_1,\xi_1)P(t_2|\alpha_2,\xi_2)
\Theta (W-|t_1-t_2|)P(\xi_1,\xi_2)dt_1dt_2}
{\int_0^{2\pi}\int_{0}^{2\pi}\int_{-\infty}^{+\infty}\int_{-\infty}^{+\infty}P(t_1|\alpha_1,\xi_1)P(t_2|\alpha_2,\xi_2)
\Theta (W-|t_1-t_2|)P(\xi_1,\xi_2)d\xi_1d\xi_2dt_1dt_2}.
\nonumber\\
\end{eqnarray}
The simple fact that $\rho(\xi_1,\xi_2|\alpha_1,\alpha_2)\ne\rho(\xi_1,\xi_2)$ brings the derivation of the original Bell (CHSH)
inequality to a halt. Indeed, in these derivations it is assumed that the probability distribution for $\xi_1$ and $\xi_2$
does not depend on the settings $\alpha_1$ or $\alpha_2$~\cite{BALL03,BELL93}.
Larsson and Gill have shown that due to the filtering by the time-coincidence window,
Eq.~(\ref{propmod3}) satisfies a modified CHSH inequality in which the upperbound of 2 is to be replaced by a number
that depends on the ratio of the number of pairs satisfying the time coincidence criterion
and the total number of pairs~\cite{LARS04}. For finite $W$, this upperbound can be larger than 2.
Hence, it is to be expected that Eq.~(\ref{propmod3}) can violate the original CHSH inequality.

By making explicit use of the time-tag model (see Eq.~(\ref{sg3b})) it can be shown that~\cite{ZHAO08} (i) if we ignore
the time-tag information ($W>T_0$), the two-particle probability takes the form of the hidden variable models
considered by Bell~\cite{BELL93}, and we cannot reproduce the results of quantum theory~\cite{BELL93},
(ii) if we focus on the case $W\rightarrow 0$ the single-particle averages are zero and the two-particle average
$E(\alpha_1 , \alpha_2)=-\cos (\alpha_1 -\alpha_2)$.

In summary, although our simulation model and its probabilistic version Eq.~(\ref{propmod2}) involve local processes only, the filtering of the
detection events by means of the time-coincidence window $W$ can produce correlations which violate Bell-type
inequalities~\cite{FINE82,PASC86,LARS04}.
Moreover, for $W\rightarrow 0$ our classical, local and causal model can produce single-particle and two-particle
averages which are the same as those of the singlet state in quantum theory.

\subsection{Violation of a Bell inequality in single-neutron interferometry}
The single-neutron interferometry experiment of Hasegawa {\it et al.}~\cite{HASE03}, demonstrating that the correlation between
the spatial and spin degree of freedom of neutrons violates a Bell-CHSH inequality, is taken to illustrate
how to construct an event-based model that reproduces this correlation by using detectors that count every neutron and without
using any post-selection procedure.

\begin{figure}[t]
\begin{center}
\includegraphics[width=10cm]{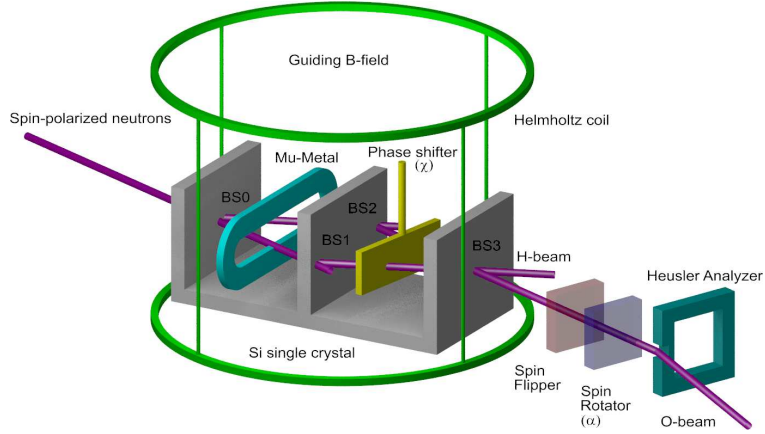}
\caption{%
Diagram of the single-neutron interferometry experiment to test a Bell inequality violation (see also Fig.~1 in Ref.~\cite{HASE03}).
}
\label{expneutron}
\end{center}
\end{figure}

A schematic picture of the single-neutron interferometry experiment is shown in Fig.~\ref{expneutron}.
Incident neutrons are passing through a magnetic-prism polarizer (not shown) that produces two spatially separated beams of
neutrons with their magnetic moments aligned parallel (spin up), respectively anti-parallel (spin down) with respect
to the magnetic axis of the polarizer which is parallel to the guiding field ${\mathbf B}$. The spin-up neutrons
impinge on a silicon-perfect-crystal interferometer~\cite{RAUC00}.
On leaving the first beam splitter, neutrons may or may not experience refraction.
A Mu-metal spin-turner changes the orientation of the magnetic moment from parallel to
perpendicular to the guiding field ${\mathbf B}$.
The result of passing through this spin-turner is that the magnetic moment of the neutrons
is rotated by $\pi/2$ ($-\pi/2$) about the $y$ axis, depending on the path followed.
Before the two paths join at the entrance plane of beam splitter BS3 a difference between the time of flights
(corresponding to a phase in the wave mechanical description) along the two paths can be manipulated by a phase shifter. The neutrons
which experience two refraction events when passing through the interferometer form the O-beam
and are analyzed by sending them through
a spin rotator and a Heusler spin analyzer.
If necessary, to induce an extra spin rotation of $\pi$, a spin flipper is placed between
the interferometer and the spin rotator. The neutrons that are selected by the Heusler spin analyzer are counted with a
neutron detector (not shown) that has a very high efficiency ($\approx 99\%$).
Note that neutrons which are not refracted by the central plate of the Si single crystal
(beam splitters BS1 and BS2) leave the interferometer
without being detected.

The single-neutron interferometry experiment yields the count rate $N(\alpha,\chi)$ for the spin-rotation angle $\alpha$ and
the difference $\chi$ of the phase shifts of the two different paths in the interferometer~\cite{HASE03}.
The correlation $E(\alpha,\chi)$ is defined by~\cite{HASE03}

\begin{equation}
E_(\alpha,\chi)=\frac{N(\alpha,\chi)+N(\alpha+\pi,\chi+\pi)-N(\alpha+\pi,\chi)-N(\alpha,\chi+\pi)}
{N(\alpha,\chi)+N(\alpha+\pi,\chi+\pi)+N(\alpha+\pi,\chi)+N(\alpha,\chi+\pi)}.
\end{equation}

If quantum theory describes the experiment it is expected that for the neutrons detected in the O-beam,
$E(\alpha,\chi)=\cos(\alpha+\chi)$ implying $|S|=E(\alpha,\chi)-E(\alpha,\chi^{\prime})+E(\alpha^{\prime},\chi)+
E(\alpha^{\prime},\chi^{\prime})>2$ for some values of $\alpha$, $\alpha^{\prime}$, $\chi$ and $\chi^{\prime}$
so that the state of the neutron cannot be written as a product
of the state of the spin and the phase. Experiments indeed show that $|S|>2$~\cite{HASE03,BART09}.

\subsubsection{Event-based model}
A minimal, discrete event simulation model of the single-neutron interferometry experiment requires a specification of the
information carried by the particles, of the algorithm that simulates the source and the interferometer components, and of the procedure
to analyze the data.

\begin{itemize}
\item{{\sl Source and particles:}
The particles (neutrons) leave the source one by one and carry a message
${\mathbf e}=(e^{i\psi_1}\cos\theta /2, e^{i\psi_2}\sin\theta /2)$
where $\psi_i=2\pi f t + \delta_i$ for $i=1,2$. Here $\delta_1 -\delta_2$ and $\theta$ specify the magnetic moment of the neutron
and $t$ specifies its time of flight where $f$ is a frequency which is characteristic for a neutron that moves with a fixed velocity $v$.
In the presence of a magnetic field ${\mathbf B}=(B_x,B_y,B_z)$, the magnetic moment rotates about the direction of ${\mathbf B}$ according
to the classical equation of motion.
}
\item{{\sl Beam splitters BS0, \ldots , BS3:}
The model for the beam splitter uses the learning process governed by the rule Eq.~(\ref{ruleofLM}).
Exploiting the similarity between the magnetic moment of the neutron and the polarization of a photon, we use the same model
for the beam splitter as the one used in Ref.~\cite{MICH11a} for polarized photons. The only difference is that we assume that
neutrons with spin up and spin down have the same reflection and transmission properties, while photons with horizontal and vertical
polarization have different reflection and transmission properties~\cite{BORN64}.
}
\item{{\sl Mu metal spin turner:}
This component rotates the magnetic moment of a neutron that follows the H-beam (O-beam) by $\pi/2$ ($-\pi/2$) about the $y$ axis.
}
\item{{\sl Spin-rotator and spin-flipper:}
The spin-rotator rotates the magnetic moment of a neutron by an angle $\alpha$ about the $x$ axis.
The spin flipper is a spin rotator with $\alpha=\pi$.
}
\item{{\sl Spin analyzer:}
This component selects neutrons with spin up, after which they are counted by a detector.
The model of this component projects the magnetic moment of the particle on the $z$ axis and sends the particle to the
detector if the projected value exceeds a pseudo-random number $r$.
}
\item{{\sl Detector:}
The detectors simply count each incoming particle, meaning that we assume that the detectors have 100\% detection efficiency.
This is an idealization of the real neutron detectors which have a detector efficiency of more than 99\%~\cite{KROU00}.
}
\end{itemize}
\begin{figure}[t]
\begin{center}
\includegraphics[width=7cm]{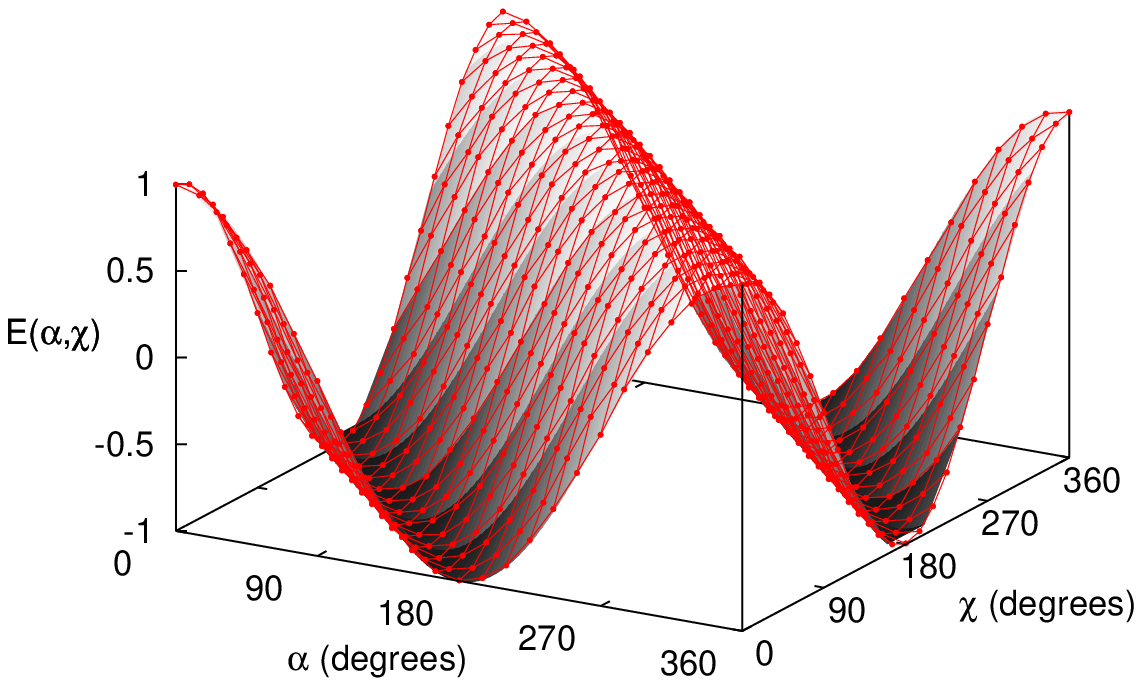}
\includegraphics[width=7cm]{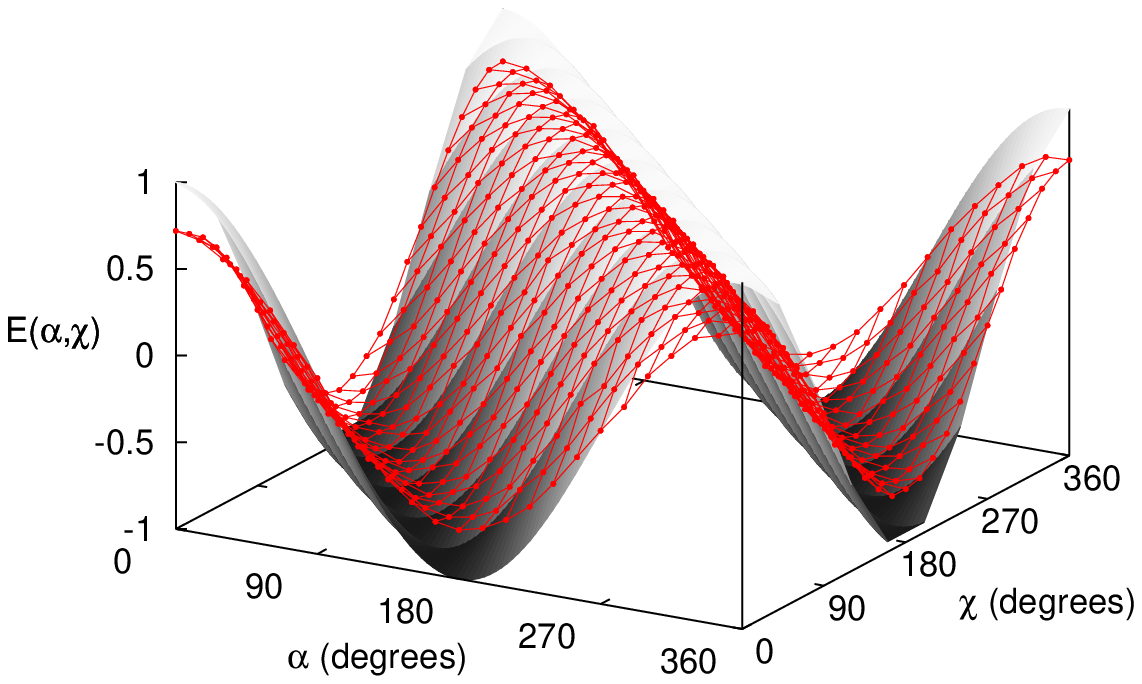}
\caption{%
Left: correlation $E(\alpha,\chi)$ between spin and path degree of freedom as obtained from an event-based simulation of the experiment
depicted in Fig.~\ref{expneutron}. Solid surface: $E(\alpha,\chi)=\cos(\alpha+\chi)$ predicted by quantum theory; circles: simulation data.
The lines connecting the markers are guides to the eye only. Model parameters: reflection percentage of BS0, \ldots, BS3 is 20\% and $\gamma=0.99$.
For each pair $(\alpha,\chi)$, four times 10000 particles were used to determine the four counts $N(\alpha,\chi)$, $N(\alpha+\pi,\chi+\pi)$,
$N(\alpha,\chi+\pi)$ and $N(\alpha+\pi,\chi+\pi)$.
Right: same as figure on the left but $\gamma =0.55$.
}
\label{simneutron}
\end{center}
\end{figure}

\subsubsection{Simulation results}
In Fig.~\ref{simneutron}(left) we present simulation results for the correlation $E(\alpha,\chi)$, assuming that the experimental conditions
are very close to ideal. For the ideal experiment quantum theory predicts that $E(\alpha,\chi)=\cos(\alpha+\chi)$.
As shown by the markers in Fig.~\ref{simneutron}, disregarding the small statistical fluctuations, there is close-to-perfect agreement
between the event-based simulation data and quantum theory.
The laboratory experiment suffers from unavoidable imperfections, leading to a reduction and distortion of the interference fringes~\cite{HASE03}.
In the event-based approach it is trivial to incorporate mechanisms for different sources of imperfections by modifying or adding update rules.
However, to reproduce the available data it is sufficient to use the parameter $\gamma$ to control the deviation from the quantum theoretical result.
For instance, for $\gamma=0.55$ the simulation (see Fig.~\ref{simneutron}) yields
$S_{max}\equiv S(\alpha=0,\chi=\pi/4,\alpha^{\prime}=\pi/2,\chi^{\prime}=\pi/4 )=2.05$, in excellent agreement with the value $2.052\pm 0.010$
obtained in experiment~\cite{HASE03}. For $\gamma=0.67$ the simulation yields $S_{max}=2.30$, in excellent agreement with the value $2.291\pm 0.008$
obtained in a similar, more recent experiment~\cite{BART09}.

\subsubsection{Working principle}
From Ref.~\cite{MICH11a} we know that the event-based model for the beam splitter produces results corresponding to those
of classical wave or quantum theory when applied in interferometry experiments.
Important for this outcome is that the phase difference $\chi$ between the two paths in the interferometer
is constant for a relatively large number of incoming particles.
If, for each incoming neutron, we pick the angle $\chi$ randomly from the same set of predetermined values to
produce Fig.~\ref{simneutron}, an event-based simulation
with $\gamma=0.99$ yields (within the usual statistical fluctuations) the correlation
$E(\alpha,\chi)\approx [\cos(\alpha+\chi)]/2$, which does not lead to a violation of the Bell-CHSH inequality (results not shown).
Thus, if the neutron interferometry experiment could be repeated with random choices for the phase shifter $\chi$ for each incident neutron,
and the experimental results would show a significant violation of the Bell-CHSH inequality, then the event-based model that we
have presented here would be ruled out.

\section{Discussion and outlook}\label{summary}

We have given a brief introduction to a new methodology for simulating,
on the level of single events, what are usually considered to be ``quantum'' phenomena.
In spirit, our approach is similar to cellular autonoma modeling advanced by Wolfram~\cite{WOLF02}
or, for instance, lattice Boltzmann modeling of fluid dynamics~\cite{SUCC01}.
The general idea is that simple rules,
which are not necessarily derived from classical Hamiltonian dynamics,
define discrete-event processes which may lead to
the (complicated) behavior that is observed in experiments.
The basic strategy in designing these rules is to carefully examine
the experimental procedure and to devise rules such that they produce the same
kind of data as those recorded in experiment, while avoiding the trap
of simulating thought experiments that are difficult to realize in the laboratory.
The \EBM\ is entirely classical in the sense that
it uses concepts of the macroscopic world and makes no reference to  \QT\
but is nonclassical in the sense that some of the rules are not those of classical Newtonian dynamics.

Depending on the experimental configuration,
the simulation of the interference and correlation phenomena, which are observed when individual photons and neutrons
are detected one by one, can make use of one or two of the four following detection processes:
\begin{itemize}
\item{(1) The detectors are simple particle counters producing a click for each incoming particle and no post-selection data procedure using a time-coincidence window is used.
The final detection efficiency is $100\%$.}
\item{(2) The detectors are simple particle counters and a post-selection data procedure using a time-coincidence window is used. Although the ratio of detected to emitted particles
is one, the final detection efficiency is less than $100\%$.}
\item{(3) The detectors are adaptive threshold detectors not producing a click for each incoming particle and no post-selection data procedure using a time-coincidence window is used.
The final detection efficiency can be less than $100\%$, depending on the experimental configuration.}
\item{(4) The detectors are adaptive threshold detectors and a post-selection data procedure using a time-coincidence window is used. The final detection efficiency is less than $100\%$.}
\end{itemize}
In Table~\ref{tab1} we give an overview of  which type of experiments our event-based approach can simulate with a particular detection process.
From Table~\ref{tab1} it can be clearly seen that
discarding detection events in a post-selection data procedure using a time-coincidence window
or using detectors which do not produce a click for each incoming particle is not a characteristic of the event-based simulation
approach for simulating interference and quantum correlation phenomena.

\begin{table*}
\caption{\label{tab1}
Overview of the event-based simulation of photon (P) and neutron (N) experiments classified according to the detection process:
(1) Detectors are simple particle counters and no post-selection
data procedure using a time-coincidence window; (2) detectors are simple particle counters and post-selection
data procedure using a time-coincidence window;
(3) detectors are adaptive threshold detectors and no post-selection
data procedure using a time-coincidence window; (4) detectors are adaptive threshold detectors and post-selection
data procedure using a time-coincidence window.
Numbers in brackets are references to the papers in which simulation results are reported.
The symbol ``V'' (``X'') indicates that it is possible (impossible) to obtain the results of quantum theory but that we did not publish the simulation results.
The symbol ``NA'' denotes that the particular detection process is not applicable to the specified experiment.
}
\begin{tabular}{lccccc}
Experiment & Particle  &(1) &  (2) & (3) & (4) \\
\hline\noalign{\smallskip}
Single beam splitter experiment                      &  P                                                      &  \cite{RAED05d,RAED05b} &   NA    &  \cite{MICH11a}         &  NA   \\
Mach-Zehnder interferometry experiment               &  P                                                      &  \cite{RAED05d,RAED05b} &   NA    &  \cite{MICH11a}         &  NA   \\
Wheeler's delayed choice experiment                  &  P                                                      &  \cite{ZHAO08b,MICH10a} &   NA    &  \cite{MICH11a}         &  NA   \\
Quantum cryptography protocols                       &  P                                                      &  \cite{ZHAO08a}         &   NA    &  V                      &  NA   \\
Quantum eraser experiment                            &  P                                                      &  \cite{JIN10c}          &   NA    &  \cite{MICH11a}         &  NA   \\
Single photon tunneling                              &  P                                                      &  V                      &   NA    &  \cite{MICH11a}         &  NA   \\
Two-beam interference experiment                     &  P                                                      &  X                      &   NA    &  \cite{JIN10b,MICH11a}  &  NA   \\
Reflection and refraction from an interface          &  P                                                      &  \cite{TRIE11}          &   NA    &  \cite{MICH11a}         &  NA   \\
Multiple beam fringes with a plane-parallel plate    &  P                                                      &  \cite{TRIE11}          &   NA    &  \cite{MICH11a}         &  NA   \\
Quantum computation                                  &  P                                                      &  \cite{RAED05c,MICH05}  &   NA    &  V                      &  NA   \\
Einstein-Podolsky-Rosen-Bohm experiment              &  P                                                      &  X                      & \cite{RAED06c,RAED07a,RAED07b,RAED07c,RAED07d,ZHAO08}& X  & \cite{MICH11a} \\
Classical correlations in HBT experiment             &  P                                                      &  X                      &   X     &  \cite{JIN10a}          &  X    \\
Quantum correlations in HBT experiment               &  P                                                      &  X                      &   V     &  X                      &   \cite{MICH11a}  \\
Interferometry                                       &  N                                                      &  Section 2.3            &   NA    &  V                      &  NA   \\
Bell-type experiment; path-spin correlations         &  N                                                      &  Section 2.3            &   NA    &  V                      &  NA    \\
Interferometry; stochastic and deterministic absorption &  N                                                   &  V                      &   NA    &  V                      &  NA   \\
Interferometry; path-spin-energy entanglement           &  N                                                   &  V                      &   NA    &  V                      &  NA   \\
Interferometry; time-dependent blocking of one path     &  N                                                   &  V                      &   NA    &  V                      &  NA    \\

\end{tabular}
\end{table*}

Three fundamental experiments, two with photons and one with neutrons, were used to illustrate our approach.
On purpose, we have kept the \EBM\ as simple as possible, perhaps
creating the impression that each experiment will require its own set of rules.
This is not the case. One universal \EBM\ for the interaction of photons
with matter suffices to explain, without altering the rules, the interference and correlation
phenomena that are observed when individual photons are detected one by one~\cite{MICH11a}.
This universal model produces the frequency distributions for observing many photons
that are in full agreement with the predictions of \MT\ and \QT~\cite{MICH11a}.
The same model was used for simulating a single neutron interferometry experiment.
Similarly, we have constructed an \EBM\ that simulates a universal quantum computer,
reproducing the probability distributions of the quantum mechanical system
without actually knowing them~\cite{MICH05}.

An important question is whether an \EBM\ leads to new predictions that may be tested experimentally.
In the stationary state (after processing many events)
the \EBM\ reproduces the statistical distributions of \QT.
Therefore, new predictions can only appear when the \EBM\ is operating in the transient regime, before the \EBM\ reaches its stationary state.
Experiments with a Mach-Zehnder interferometer consisting of two independent $50\%$ reflective beam splitters and a phase shifter (not an integrated Mach-Zehnder interferometer)
and two-beam interference that may be able
to address this issue have been discussed in Ref.~\cite{MICH12a} and Ref.~\cite{JIN10b}, respectively.
We hope that our simulation work will stimulate the design of new experiments
to test the applicability of our approach to event-based processes and to exclude some of our non-unique models.

Finally, it may be of interest to mention that the discrete-event approach
reviewed in this paper may open a route to
rigorously include the effects of interference in ray-tracing software.
For this purpose, it is necessary to extend the \EBM\ to include diffraction, evanescent waves
and the behavior of the Lorentz model for the response of material
to the electromagnetic field~\cite{TAFL05}.
We leave these extensions for future research.

\begin{acknowledgement}
We would like to thank K. De Raedt, F. Jin, and S. Miyashita for many thoughtful comments
and contributions to the work on which this review is based.
This work is partially supported by NCF, the Netherlands.
\end{acknowledgement}

\bibliographystyle{adp}
\bibliography{c:/d/papers/epr11,c:/d/papers/neutrons}

\end{document}